# Analysis of some properties for a basic Petri net model

**Assist.Prof. Alexandra Fortiş, PhD Candidate**
**"Tibiscus" University, Timişoara**

REZUMAT. Formalismul modelelor cu reţele Petri oferă o bază teoretică solidă, susţinută de metode matematice puternice, capabile să extragă informaţii necesare pentru formalismul şi simularea sistemelor reale care prezintă caracteristici de concurenţă şi sincronizare. Lucrarea prezintă un model bazat pe o reţea Petri, în scopul extragerii unor informaţii relativ la procesul tehnologic de producere a unui aditivul alimentar.

## 1 Introduction

The target of this research is to provide a novel approach for materials balances modeling using Petri nets, in order to provide a model easily understandable by both business analysts and technological engineers. The natural way to complete this task seems to be through symbolic computation.

Materials and energy balances have a high level of significance for food industry and their theory is based on the fundamental laws of conservation. There are several models that can be used for materials and energy balances. These models are CAPE-based, the Computer-Aided Production Engineering being concerned with:
- design and engineering of a system;
- standard engineering methods;
- problem-solving techniques.

Several models can be identified arising from different points of view, but the goal of this paper is to present a Petri net-based model, which can be used for identification of basic properties of the underlying workflow system, in order to extract theoretical result.

## 2 Technological background

In this paper we are going to provide a Petri net based model for the technological flow of baking soda production. Below we present the block schema for the processes involved in. The process is composed of seven sub-processes. Briefly, we start with calcium carbonate (limestone) which is heated. An independent sub-process is to dissolve ammonia in brine. The ammoniated brined is then bubbled with $CO_2$. Follow the moment when the independent sub-processes interfere.

$NaCl + NH_3 + H_2O + CO_2$ 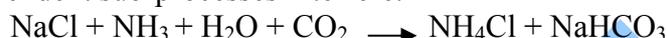 $NH_4Cl + NaHCO_3$

The obtained substances are cooled down to $0^0C$ so that $NaHCO_3$ precipitates out. The filtered $NaHCO_3$ is heated in order to obtain $Na_2CO_3$, process called *decarbonation*. Another product of the first reaction, CaO, is dissolved in water. Finally, the filtrate product $NH_4Cl$ is reacted with $Ca(OH)_2$ in order to obtain the desired product.

## 3 Theoretical aspects related to Petri net model

Petri nets were chosen as modeling tool because their formalism offers a strong theoretical base, sustained by powerful mathematical methods for the purpose of modeling and analysis of operation-flow processes. At this point, an analysis of the net properties can be performed, in order to underline properties such as *safeness*, *boundedness*, *conservation*, *liveness*. Also, invariants can be determined, this information being used in order to analyse the marking of the Petri net model and the conservation of tokens. This is where symbolic computation could be needed.

Recall first the definition for a Petri net:

**Definition**. A Petri net is a tuple N = (P, T, F) with
- P, T two disjoint sets, P for places and T for transitions
- F a function, $F \subseteq P \times T \cup T \times P$ representing the net flow, designating the connections between the places and the transitions.

For our example, the Petri net is completely defined by the following set of data:

$$P = \{P0, P1, P2,\ldots, P18\}$$
$$T = \{T0, T1,\ldots, T7\}$$
$$F=\{(P0,T0), (T0,P3), (T0,P4), (P6,T2), (P1,T1), (P2,T1),$$
$$(T1,P5), (T2,P7), (T2,P8), (P10,T4), (T4,P11), (P7,T5),$$
$$(P11,T5), (P8,T6), (T5, P12), (T5, P13), (T5, P14),$$
$$(T6,P15), (P15,T7), (T7, P16), (T7,P17), (T7, P18)\}$$

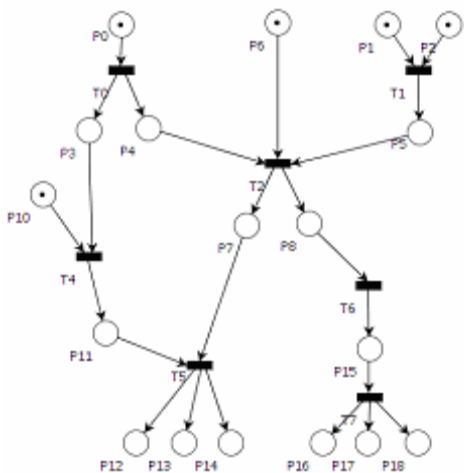

**Remark**. At this stage one has to notice that in order to produce a model for a specified process we have to:
- establish the static structure of the system;
- establish rules for the dynamic behavior of the system.

Firstly we have designated places, transitions and interconnectivity between them. Practically, such kind of model is, in fact, presenting a static structure of the system.

In order to obtain further information, we are stating that the elements of the set P are locations (variables) and the elements of the set T are actions. In this context, the state of the system can be interpreted as the content of locations at a certain moment. In our example places represent chemical substances and transitions are equivalent to chemical processes. Because we present only a simple net-based model, we are using tokens only for pointing out which are the starting points for each sub-process involved in this analysis. Adding tokens to the starting places is an important procedure, because in this way one knows that at the actual time, the system can produce.

19

The net presented here is with *asynchronous* execution meaning that some of the transitions (ex. T0 and T1, T2 and T4, T5 and T6) can be fired at the same moment. By the other hand, transition T2 can be fired only after the completion of T0 and T1, T4 and T6 only after T2, T5 after T2 and T4 and T7 after T6, so there is also a *synchronous* aspect for this net.

We can develop our model by means of linear algebra techniques and we are using for this analysis the tool PIPE2[1]: Platform Independent Petri net Editor 2.4 for Windows.

**4 Net analysis results**

**4.1 Petri net invariant analysis results**

The P-Invariants associated with this Petri net are presented below

```
0 0 0 0 0 0 0 1 0 0 1 0 1 1 1 0 1 0 0 1 0 1 0 1 0 0 0
0 0 0 1 1 1 0 0 0 0 0 0 0 0 0 0 0 1 0 1 0 1 0 0 0 0
1 1 1 0 0 0 0 0 0 0 0 0 0 0 0 0 0 0 0 0 0 0 0 0 0 0 0
1 1 1 0 0 0 0 0 0 0 0 0 0 0 1 1 0 1 0 0 0 0 0 0 0 0 0
1 0 0 0 0 0 0 0 0 0 0 0 0 1 0 0 0 0 1 1 0 0 0 0 1 0 0
0 1 0 0 0 0 0 0 0 0 0 0 0 0 1 1 0 0 0 0 1 1 0 0 0 1 0
0 0 1 0 0 0 0 0 0 0 0 0 0 0 0 0 1 1 0 0 0 0 1 1 0 0 1
0 0 0 1 1 1 1 1 1 1 1 1 0 0 0 0 0 0 0 0 0 0 0 0 0 0 0
0 0 0 1 0 0 1 0 0 0 0 0 1 0 0 0 0 0 0 0 0 0 0 0 0 0 0
0 0 0 0 1 0 0 1 1 1 0 0 0 0 0 0 0 0 0 0 0 0 0 0 0 0 0
0 0 0 0 0 1 0 0 0 0 1 1 1 0 0 0 0 0 0 0 0 0 0 0 0 0 0
0 0 0 0 0 0 1 0 0 1 0 0 0 0 0 0 0 0 0 0 0 0 0 0 1 1 1
0 0 0 0 0 0 0 0 0 0 0 0 0 1 1 0 1 0 0 0 0 0 0 0 0 0 0
0 0 0 0 0 0 0 1 0 0 1 0 1 0 0 0 0 0 0 1 0 1 0 1 0 0 0
0 0 0 1 1 1 0 1 0 0 1 0 0 0 0 0 0 0 1 0 1 0 1 0 1 1 1
0 0 0 0 0 0 1 0 0 1 0 0 1 0 0 0 1 0 1 0 0 0 0 0 0 0 0
0 0 0 0 0 0 0 0 0 0 0 0 0 0 0 0 1 0 1 1 1 1 1 1 1 1 1
0 0 0 1 1 1 1 1 1 1 1 1 1 0 0 0 0 0 0 0 0 0 0 0 0 0 0
```

---

[1] PIPE2: http://pipe2.sourceforge.net/

Since the Petri net is covered by positive P-Invariants, it is bounded. The associated P-Invariant equations are presented below:

$$M(P10) + M(P11) + M(P12) = 1$$
$$M(P10) + M(P11) + M(P13) = 1$$
$$M(P10) + M(P11) + M(P14) = 1$$
$$M(P1) + M(P15) + M(P16) + M(P5) + M(P8) = 1$$
$$M(P1) + M(P15) + M(P17) + M(P5) + M(P8) = 1$$
$$M(P1) + M(P15) + M(P18) + M(P5) + M(P8) = 1$$
$$M(P15) + M(P16) + M(P6) + M(P8) = 1$$
$$M(P15) + M(P17) + M(P2) + M(P5) + M(P8) = 1$$
$$M(P0) + M(P15) + M(P17) + M(P4) + M(P8) = 1$$
$$M(P15) + M(P17) + M(P6) + M(P8) = 1$$
$$M(P15) + M(P18) + M(P2) + M(P5) + M(P8) = 1$$
$$M(P0) + M(P15) + M(P18) + M(P4) + M(P8) = 1$$
$$M(P15) + M(P18) + M(P6) + M(P8) = 1$$
$$M(P0) + M(P15) + M(P16) + M(P4) + M(P8) = 1$$
$$M(P0) + M(P11) + M(P12) + M(P3) = 1$$
$$M(P0) + M(P11) + M(P13) + M(P3) = 1$$
$$M(P13) + M(P6) + M(P7) = 1$$
$$M(P0) + M(P11) + M(P14) + M(P3) = 1$$
$$M(P14) + M(P6) + M(P7) = 1$$
$$M(P1) + M(P12) + M(P5) + M(P7) = 1$$
$$M(P0) + M(P12) + M(P4) + M(P7) = 1$$
$$M(P1) + M(P13) + M(P5) + M(P7) = 1$$
$$M(P0) + M(P13) + M(P4) + M(P7) = 1$$
$$M(P1) + M(P14) + M(P5) + M(P7) = 1$$
$$M(P0) + M(P14) + M(P4) + M(P7) = 1$$
$$M(P12) + M(P2) + M(P5) + M(P7) = 1$$
$$M(P13) + M(P2) + M(P5) + M(P7) = 1$$
$$M(P14) + M(P2) + M(P5) + M(P7) = 1$$

## 4.2 Petri net incidence and marking

The forwards incidence matrix ($I^+$), backwards incidence matrix ($I^-$), and combined incidence matrix ($I$) are presented in the following tables:

21

**Table 1**

| Forwards incidence matrix $I^+$ | | | | | | | | Backwards incidence matrix $I^-$ | | | | | | | | Combined incidence matrix $I$ | | | | | | | |
|---|---|---|---|---|---|---|---|---|---|---|---|---|---|---|---|---|---|---|---|---|---|---|---|
| | T0 | T1 | T2 | T4 | T5 | T6 | T7 | | T0 | T1 | T2 | T4 | T5 | T6 | T7 | | T0 | T1 | T2 | T4 | T5 | T6 | T7 |
| P0 | 0 | 0 | 0 | 0 | 0 | 0 | 0 | P0 | 1 | 0 | 0 | 0 | 0 | 0 | 0 | P0 | -1 | 0 | 0 | 0 | 0 | 0 | 0 |
| P1 | 0 | 0 | 0 | 0 | 0 | 0 | 0 | P1 | 0 | 1 | 0 | 0 | 0 | 0 | 0 | P1 | 0 | -1 | 0 | 0 | 0 | 0 | 0 |
| P10 | 0 | 0 | 0 | 0 | 0 | 0 | 0 | P10 | 0 | 0 | 1 | 0 | 0 | 0 | 0 | P10 | 0 | 0 | 0 | -1 | 0 | 0 | 0 |
| P11 | 0 | 0 | 0 | 1 | 0 | 0 | 0 | P11 | 0 | 0 | 0 | 0 | 1 | 0 | 0 | P11 | 0 | 0 | 0 | 1 | -1 | 0 | 0 |
| P12 | 0 | 0 | 0 | 0 | 1 | 0 | 0 | P12 | 0 | 0 | 0 | 0 | 0 | 0 | 0 | P12 | 0 | 0 | 0 | 0 | 1 | 0 | 0 |
| P13 | 0 | 0 | 0 | 0 | 1 | 0 | 0 | P13 | 0 | 0 | 0 | 0 | 0 | 0 | 0 | P13 | 0 | 0 | 0 | 0 | 1 | 0 | 0 |
| P14 | 0 | 0 | 0 | 0 | 1 | 0 | 0 | P14 | 0 | 0 | 0 | 0 | 0 | 0 | 0 | P14 | 0 | 0 | 0 | 0 | 1 | 0 | 0 |
| P15 | 0 | 0 | 0 | 0 | 0 | 1 | 0 | P15 | 0 | 0 | 0 | 0 | 0 | 0 | 1 | P15 | 0 | 0 | 0 | 0 | 0 | 1 | -1 |
| P16 | 0 | 0 | 0 | 0 | 0 | 0 | 1 | P16 | 0 | 0 | 0 | 0 | 0 | 0 | 0 | P16 | 0 | 0 | 0 | 0 | 0 | 0 | 1 |
| P17 | 0 | 0 | 0 | 0 | 0 | 0 | 1 | P17 | 0 | 0 | 0 | 0 | 0 | 0 | 0 | P17 | 0 | 0 | 0 | 0 | 0 | 0 | 1 |
| P18 | 0 | 0 | 0 | 0 | 0 | 0 | 1 | P18 | 0 | 0 | 0 | 0 | 0 | 0 | 0 | P18 | 0 | 0 | 0 | 0 | 0 | 0 | 1 |
| P2 | 0 | 0 | 0 | 0 | 0 | 0 | 0 | P2 | 0 | 1 | 0 | 0 | 0 | 0 | 0 | P2 | 0 | -1 | 0 | 0 | 0 | 0 | 0 |
| P3 | 1 | 0 | 0 | 0 | 0 | 0 | 0 | P3 | 0 | 0 | 0 | 1 | 0 | 0 | 0 | P3 | 1 | 0 | 0 | -1 | 0 | 0 | 0 |
| P4 | 1 | 0 | 0 | 0 | 0 | 0 | 0 | P4 | 0 | 0 | 1 | 0 | 0 | 0 | 0 | P4 | 1 | 0 | -1 | 0 | 0 | 0 | 0 |
| P5 | 0 | 1 | 0 | 0 | 0 | 0 | 0 | P5 | 0 | 0 | 1 | 0 | 0 | 0 | 0 | P5 | 0 | 1 | -1 | 0 | 0 | 0 | 0 |
| P6 | 0 | 0 | 0 | 0 | 0 | 0 | 0 | P6 | 0 | 0 | 1 | 0 | 0 | 0 | 0 | P6 | 0 | 0 | -1 | 0 | 0 | 0 | 0 |
| P7 | 0 | 0 | 1 | 0 | 0 | 0 | 0 | P7 | 0 | 0 | 0 | 0 | 1 | 0 | 0 | P7 | 0 | 0 | 1 | 0 | -1 | 0 | 0 |
| P8 | 0 | 0 | 1 | 0 | 0 | 0 | 0 | P8 | 0 | 0 | 0 | 0 | 0 | 1 | 0 | P8 | 0 | 0 | 1 | 0 | 0 | -1 | 0 |

The initial markings associated with this Petri net are presented in the following Figure:

Marking

| | P0 | P1 | P10 | P11 | P12 | P13 | P14 | P15 | P16 | P17 | P18 | P2 | P3 | P4 | P5 | P6 | P7 | P8 |
|---|---|---|---|---|---|---|---|---|---|---|---|---|---|---|---|---|---|---|
| Initial | 1 | 1 | 1 | 0 | 0 | 0 | 0 | 0 | 0 | 0 | 0 | 1 | 0 | 0 | 0 | 1 | 0 | 0 |
| Current | 1 | 1 | 1 | 0 | 0 | 0 | 0 | 0 | 0 | 0 | 0 | 1 | 0 | 0 | 0 | 1 | 0 | 0 |

Finally, the information about enabled transitions is specified in the following Figure:

Enabled transitions

| T0 | T1 | T2 | T4 | T5 | T6 | T7 |
|---|---|---|---|---|---|---|
| yes | yes | no | no | no | no | no |

### 4.3 Reachability Graph Results

Even if not shown in a separate Figure, here is the evaluation of the reachability graph: PIPE2 needed 0.781seconds for the reachability graph; while converting to dot format and constructing took 3.235 seconds, and the total time for analysis was 4.016 seconds.

### 4.4 Petri net state space analysis results

Finally, the following results synthesize the information contained in our target Petri net.
The result for the boundedness problem: the target Petri net is bounded.
The result for the safeness problem: the target Petri net is safe.
The result for the deadlock problem: there is a possibility of deadlock, illustrated by the shortest path to deadlock: T0 T1 T2 T4 T5 T6 T7

### Conclusions and future work

As presented before, a Petri net model is a graphical tool which could become a communication path between theoreticians and practitioners, being suitable to concurrent, asynchronous, distributed, parallel, nondeterministic and/or stochastic real systems. Petri nets provide also a design language for specification of workflows and analysis techniques to verify the workflow procedure, aspects which we are intended to include in future work.